\documentclass[conference]{IEEEtran}
\IEEEoverridecommandlockouts

\usepackage{cite}
\usepackage{amsmath,amssymb,amsfonts}
\usepackage{algorithmic}
\usepackage{graphicx}
\usepackage{textcomp}
\usepackage{xcolor}
\usepackage{subcaption}
\usepackage{amssymb}
\def\BibTeX{{\rm B\kern-.05em{\sc i\kern-.025em b}\kern-.08em
    T\kern-.1667em\lower.7ex\hbox{E}\kern-.125emX}}
\begin{document}

\title{Dynamic Connectivity and Local Frequency Strength under Stochastic Variations\\
\thanks{This work was supported by São Paulo Research
Foundation (FAPESP) under grant numbers 25/12222-5 and 21/11380-5.}
}

 \author{
\IEEEauthorblockN{Bruno Pinheiro and Daniel Dotta}
\IEEEauthorblockA{University of Campinas (UNICAMP)\\
Campinas, Brazil \\ b229989@dac.unicamp.br,dottad@unicamp.br}
}

\maketitle

\begin{abstract}
This paper introduces a novel metric, termed the Generalized Fiedler Vector (GFV), to evaluate the \textit{dynamic connectivity} in power systems. The proposed metric leverages the network connectivity, represented by the system Laplacian matrix, together with the nodal inertia distribution, following a formulation previously developed by the first author. By capturing the interplay between system topology and dynamic properties, the GFV provides valuable insights for the optimal siting of stochastic generation to mitigate its impact on local and system-wide frequency variability. The effectiveness of the proposed approach is demonstrated through Monte Carlo simulations performed on the IEEE 68-bus test system.
\end{abstract}


\section{Introduction}

\subsection{Motivation}

The increasing integration of inverter-based resources (IBRs) poses significant challenges for power system planning and operation. The stochastic feed-in of IBRs during normal operating conditions induces frequency variations that propagate across the network, influencing the dynamic behavior of other devices \cite{Hantao2023}. The spatial placement of IBRs directly affects both the magnitude and distribution of these fluctuations \cite{bruno_SIN}. Large ambient frequency variations can accelerate mechanical wear in synchronous machines, increase local control effort in IBRs \cite{Misyris2022}, and even trigger the maloperation of protection systems \cite{TROVATO2019}. Understanding how IBR placement and system topology jointly shape frequency dynamics is therefore essential for ensuring reliable operation in low-inertia grids. Building upon recent work in \cite{effecInertia}, this paper introduces a novel metric that integrates network topology and the spatial distribution of inertia to quantify dynamic connectivity, offering new insights into local system strength.

\subsection{Literature Review}

Recent studies have investigated the stochastic impact of renewable generation on frequency dynamics. In \cite{Zuo2021}, the interaction between grid-forming and grid-following battery energy storage systems (BESS) under stochastic disturbances was analyzed, showing the superior frequency control of grid-forming units, although without addressing locational or topological effects. In \cite{Vaca2025}, power fluctuation impacts on frequency variability were assessed through an analytical expression of the center-of-inertia (COI) frequency, emphasizing the influence of system topology on frequency statistics.

The role of network structure has also been examined in \cite{Warren_EPSR2021}, where the influence of topology on COI frequency and rate of change of frequency (RoCoF) was quantified using a network-based metric, revealing the sensitivity of system-wide frequency response to the spatial distribution of stochastic sources. Similarly, \cite{Jade2024} introduced two novel indices, the \textit{inertia clustering coefficient} and the \textit{inertia centrality coefficient}, to evaluate the spatial distribution of inertia by accounting for both network connectivity and control parameter allocation. These metrics demonstrated a strong correlation between inertia dispersion and synchronization performance following disturbances. However, the locational effects on nodal frequency behavior and the role of inertia within a spatially dependent framework remain largely unexplored. In \cite{Brahma2020}, the concept of dynamic flexibility was introduced via a sensitivity-based analytical index that quantifies a system’s capability to withstand local generation or load fluctuations, primarily addressing small-signal stability. An inertial index derived from oscillation observability was also proposed to complement this perspective. Similarly, \cite{Balakrushna2023} investigated the impact of IBR placement using an inertia-based metric, highlighting the link between inertia distribution and oscillatory behavior. However, most existing metrics rely on sensitivity analyses or COI-based measures, which limits both their spatial resolution and general applicability.

\subsection{Contributions \& Organization}

This paper proposes a novel analytical metric, termed the Generalized Fiedler Vector (GFV), which integrates the network Laplacian and the nodal inertia distribution to characterize \textit{dynamic connectivity}, interpreted as a measure of nodal frequency strength. The GFV identifies nodes more resilient to stochastic power fluctuations, such as those arising from variable renewable generation or load variability. The proposed formulation provides a locationally sensitive perspective of frequency stability, bridging the gap between network connectivity and inertial support. 

The remainder of this paper is organized as follows. Section II reviews the nodal inertia distribution and disturbance propagation problem. Section III presents the proposed methodology. Section IV discusses the simulation setup and results, and Section V concludes the paper.

\section{Background}

\subsection{Network Laplacian Matrix}

A power system with $n_b$ load buses and $n_g$ generators can be represented using graph theory, allowing the use of network attributes to describe its dynamic behavior. In this work, the network Laplacian matrix is derived from the linearized power flow model:
\begin{equation}
    \label{eq:DCpowerflow}
    \Delta \boldsymbol{P_e} =  \boldsymbol{L} \Delta \boldsymbol{\theta},
\end{equation}

\noindent where $\Delta \boldsymbol{P_e}\in\mathbb{R}^{n}$ with $n = n_b + n_g$ is the vector of active power variations; $\Delta \boldsymbol{\theta}\in\mathbb{R}^{n}$ represents bus angle variations; and $\boldsymbol{L}\in\mathbb{R}^{n \times n}$ is the Laplacian matrix, computed as:
\begin{equation}
\label{eq:Laplacian}
    \boldsymbol{L}_{i,j} = 
        \begin{cases}
         -|V^0_i||V^0_j|B_{i,j}\cos{(\theta_i^0 - \theta_j^0)}, & \text{if } i \neq j \\
        \sum_{k, k \neq i} |V^0_i||V^0_k|B_{i,k}\cos{(\theta_i^0 - \theta_k^0)}, & \text{if } i = j \\
\end{cases} 
\end{equation}
\noindent where $|V^0_i|$ and $\theta^0_i$ denote the voltage magnitude and angle at bus $i$, and $B_{i,j}$ is the susceptance between buses $i$ and $j$. Conductances are neglected. The off-diagonal element $L_{i,j}$ represents the synchronizing power coefficient between nodes $i$ and $j$.

The resulting Laplacian matrix is symmetric and positive semi-definite, capturing the connectivity of the electrical network. It has one zero eigenvalue, while the remaining eigenvalues satisfy $0=\lambda_1 < \lambda_2 < \lambda_3 < \dots < \lambda_n$. The eigenpairs $(\lambda_\alpha, \boldsymbol{\phi}_\alpha)$ are obtained from $\boldsymbol{L}\boldsymbol{\phi}_\alpha = \lambda_\alpha\boldsymbol{\phi}_\alpha$. 

The second smallest eigenvalue, $\lambda_2$, is known as the \textit{algebraic connectivity} and quantifies the overall connectedness of the network: smaller $\lambda_2$ indicates weaker connectivity or the presence of a bottleneck. The corresponding eigenvector, $\boldsymbol{\phi}_2$, is called the \textit{Fiedler vector} and reveals the nodes that form the weakest connection between two “subsystems” in the network.

\subsubsection{Spectral Analysis and Frequency Variability}

As shown in~\cite{Pagnier2019}, the Fiedler eigenvector of the reduced Laplacian matrix $\boldsymbol{L}_{\textrm{red}}$ strongly influences the spatial pattern of frequency variations following a disturbance. In a homogeneous system, where all inertia constants $m$ and damping coefficients $d$ are identical, a disturbance at bus $b$ produces a frequency deviation at bus $i$ that can be expressed by the spectral decomposition of the Laplacian matrix~\cite{Pagnier2019}:
\begin{equation}
    \Delta f_i(t) = \frac{\Delta P e^{\frac{-\gamma t}{2}}}{m} \sum_{\alpha = 1}^{n} \phi_{\alpha i}\phi_{\alpha b} \frac{\sin{\left(\sqrt{\lambda_{\alpha}/m - \gamma^2 / 4}t\right)}}{\sqrt{\lambda_{\alpha}/m - \gamma^2 / 4}},
\end{equation}
\noindent where $\Delta f_i(t)$ is the frequency deviation at bus $i$; $\gamma = d/m$; $\phi_{\alpha i}$ is the $i$-th element of the $\alpha$-th eigenvector; and $\lambda_{\alpha}$ is the corresponding eigenvalue.
This formulation reveals that the frequency response of a bus depends on both the eigenstructure of $\boldsymbol{L}_{\textrm{red}}$ and the location of the disturbance. Buses associated with larger absolute values in the Fiedler vector tend to exhibit greater frequency deviations, while smaller values indicate more robust, disturbance-resistant locations.

Therefore, the Fiedler vector serves as a useful indicator for identifying critical or vulnerable nodes in the system. However, its traditional formulation assumes homogeneous inertia and neglects nodal inertia heterogeneity, limiting its applicability to realistic, low-inertia power systems.

\subsection{Nodal Inertia Matrix}

The local rate of change of frequency (RoCoF) at a bus depends on system topology, operating conditions, and the inertial response of synchronized generators \cite{Bruno2023}. After a disturbance at bus $j$, the local RoCoF can be analytically related to the nodal inertia, which represents the cumulative contribution of all generators to the local frequency response.

An analytical formulation for nodal inertia, derived without the need for dynamic simulations, was proposed in~\cite{Bruno2023}. Based on generator inertia constants, operating points, and equivalent susceptances, the nodal inertia at bus $j$ is given by:
\begin{equation}
\label{eq:hj}
h_{j} = \frac{\sum_{k = 1}^{n_{g}} B_{k,j}E_{k}\cos(\delta_{k0} - \theta_{j0})}{\sum\limits_{i = 1}^{n_{g}} H_{i}^{-1}D_{j,i}B_{i,j} E_{i}\cos (\delta_{i0} - \theta_{j0})},
\end{equation}

\noindent where $E_k$ is the internal voltage of generator $k$; $\delta_{k0}$ and $\theta_{j0}$ are the generator and bus voltage angles at the operating point; $H_i$ is the inertia constant of generator $i$; $D_{j,i}$ is the $(j,i)$ element of the frequency participation matrix \cite{Milano_FDF}:
\begin{equation}
\label{eq:D_matrix}
\textbf{D} = -\textbf{B}_{\text{ext}}^{-1}\textbf{B}_{g}
\end{equation}

\noindent and $B_{k,j} = \text{Im}{\textbf{Y}^{\text{red}}}_{k,j}$ represents the equivalent susceptance obtained from the reduced admittance matrix:
\begin{equation}
\label{eq:transfer_matrix_2}
\textbf{Y}^{\text{red}}_{j} = \textbf{Y}_{B_jB_j} - \textbf{Y}_{G_jB_j}\textbf{Y}_{B_jB_j}^{-1}\textbf{Y}_{B_jG_j}.
\end{equation}

Unlike the Fiedler vector, which captures modal behavior and disturbance propagation patterns, nodal inertia focuses on the immediate inertial response following a disturbance. However, it neglects subsequent oscillatory interactions governed by network dynamics. To overcome these limitations, the next section introduces the dynamic connectivity metric, which integrates the spatial and inertial characteristics of the system into a unified framework.

\section{Proposed Generalized Fiedler Vector}

Traditional spectral methods based on the classical Fiedler vector describe how disturbances propagate through the network purely as a function of topology. However, these approaches implicitly assume homogeneous dynamical behavior across buses and therefore neglect the spatial variability of inertia and the influence of load buses. To address this limitation, we introduce the notion of \textit{dynamic connectivity}, which generalizes algebraic connectivity by embedding nodal inertia into the spectral representation of the system.

The construction of dynamic connectivity begins with the system operating point and network topology. From these, the standard Laplacian matrix $\mathbf{L}$ is obtained using~\eqref{eq:Laplacian}. Next, the nodal inertia values $h_j$ are computed for all buses according to~\eqref{eq:hj}, allowing the formation of the diagonal inertia matrix $\mathbf{N}\in\mathbb{R}^{n_b\times n_b}$ defined as
\begin{equation}
    \textbf{N} = \text{diag}\{h_1,h_2,...,h_{n_b}\}.
\end{equation}
Dynamic connectivity is then characterized through the generalized eigenvalue problem (GEP):
\begin{equation}
\label{eq:GEP}
    \textbf{L}\bar{\phi}_i = \bar{\lambda_i} \textbf{N}\bar{\phi}_i,
\end{equation}
where $\bar{\lambda}_i$ and $\bar{\phi}_i$ denote the $i$th generalized eigenvalue and eigenvector of the matrix pencil $(\mathbf{L},\mathbf{N})$, respectively. This formulation naturally integrates inertia into the network’s spectral properties, effectively weighting the Laplacian according to local dynamic characteristics.

\vspace{0.15cm}
\noindent \textbf{Definition (Generalized Fiedler Vector – GFV).}
The GFV is defined as the normalized right eigenvector associated with the smallest nonzero eigenvalue of the generalized eigenvalue problem in~\eqref{eq:GEP}:
\begin{equation}
\label{eq:GFV}
\textbf{GFV} = \frac{\boldsymbol{|\bar{\phi}}_2|}{\max(\boldsymbol{|\bar{\phi}}_2|)}.
\end{equation}

\noindent \textbf{Remark 1}: The associated eigenvalue $\bar{\lambda}_2$ represents the \emph{dynamic connectivity} of the system, while the GFV encodes how this connectivity varies spatially.

\noindent \textbf{Definition (Nodal frequency strength).} Nodal frequency strength is defined as the ability of an individual bus to locally resist, absorb, and limit the propagation of frequency deviations arising from power imbalances. It reflects how resilient a node is to stochastic or localized disturbances, considering both network topology and inertia distribution. 

\noindent \textbf{Remark 2}: Buses with lower GFV values exhibit greater \textit{nodal frequency strength}, meaning they are more resilient to local or stochastic power fluctuations, whereas buses with higher values are more sensitive to frequency disturbances and play a stronger role in their propagation. Thus, the GFV provides a locational, inertia-aware indicator of frequency resilience and offers a physically interpretable link between network topology, inertia distribution, and disturbance propagation.

\section{Case Study}

This section applies the proposed methodology to determine the nodal frequency strength in the IEEE 68-bus system. This system consists of 16 synchronous generators (SG). In the simulations, all machines are represented by a sixth-order model. The detailed parameters of the generators and controllers can be found in~\cite{Benchmark}. The analysis focuses on locational frequency variability rather than frequency nadir or long-term frequency recovery, and therefore emphasizes spatial frequency behavior prior to dominant turbine-governor action. The results were validated through 1000 Monte Carlo simulations conducted using the PST (Power System Toolbox) in Matlab~\cite{PST}, where local and global frequency variations (COI), and their propagation throughout the system were analyzed. 

\vspace{-0.1cm}
\subsection{Analytical Results}

\begin{figure*}[htb]
\includegraphics[width=7.0in]{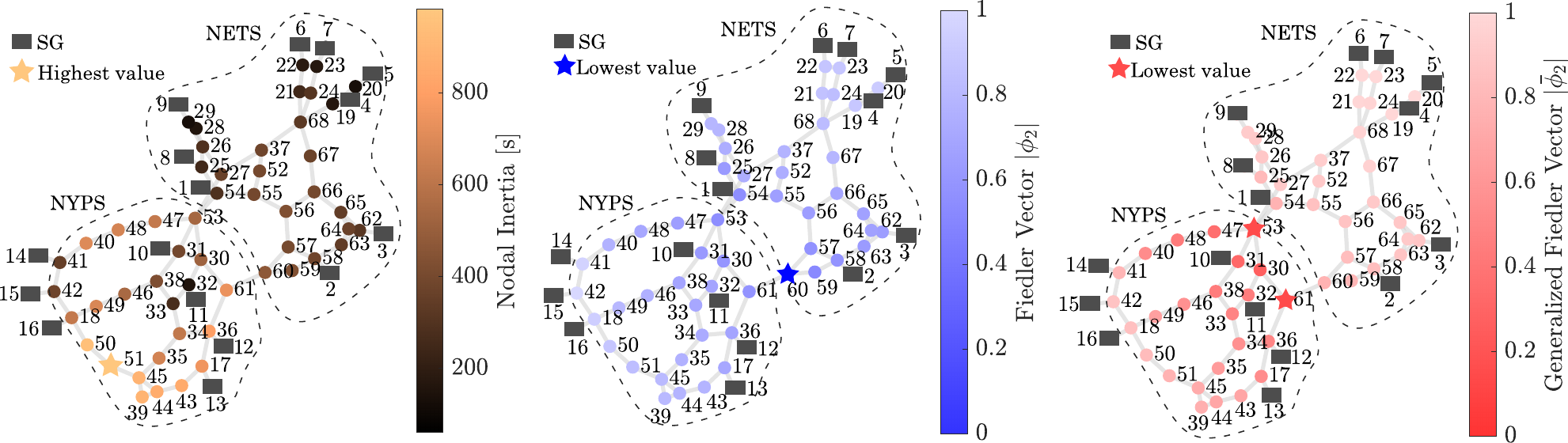}
\centering
\caption{(a) nodal inertia, (b) Fiedler eigenvector, and (c) Generalized Fiedler eigenvector (GFV).}
\label{fig:all}
\end{figure*}

Figure~\ref{fig:all} compares the spatial distribution of nodal inertia, the Fiedler eigenvector, and the proposed Generalized Fiedler Vector (GFV) for the IEEE 68-bus system. In Fig.~\ref{fig:all}a, buses with high inertia are concentrated in the NYPS area (17, 36, 39, 43–45, 50, 51, and 61), with Bus~51 presenting the largest value. Conversely, buses~20 and~29 exhibit the lowest inertia, indicating higher expected RoCoF under local contingencies. Figure~\ref{fig:all}b shows the Fiedler eigenvector, which reflects the system’s algebraic connectivity. Low-magnitude elements (buses~54, 59, and~60 in the NETS area) correspond to well-connected nodes, while large magnitudes at peripheral buses (5–7, 9, 14–16) reveal weak coupling and potential inter-area boundaries. 

Fig.~\ref{fig:all}c presents the proposed GFV, which extends the classical definition by incorporating the distribution of nodal inertia into the Laplacian formulation. In this representation, large absolute values of GFV correspond to low nodal frequency strength, whereas small absolute values indicate stronger dynamic coupling. Notably, buses~53, 30, and~61 (in red) exhibit the smallest values of GFV, suggesting that these locations possess the greatest nodal frequency strength. Conversely, the light red-colored buses, corresponding primarily to the NETS and parts of the NYPS areas, reveal lower frequency strength. It should be noted that, compared to Fiedler vector, the proposed GFV clearly demonstrates that nodes within the NYPS region and on the boundary with the NETS region are more robust.

An important observation arises when comparing Figs.~\ref{fig:all}a and~\ref{fig:all}c: buses with high inertia (e.g.,~50, 51, 45, and~39) are not necessarily those with the highest frequency strength according to the GFV. This discrepancy emphasizes that the proposed metric captures not only the amount of local inertia but also how inertia is spatially distributed and coupled through the network. In this sense, the GFV thus provides a broader notion of \textit{dynamic connectivity}, capturing how both network topology and inertia distribution shape the system’s frequency response.

\vspace{-0.1cm}
\subsection{Monte Carlo Validation}

Based on the obtained results, we select four representative buses to assess the impact of a stochastic feed-in generation from a IBR in buses with different nodal inertia levels. In this sense, an increase of 500~MW of demand are uniformly distributed in all system buses and a Type-3 double feed induction generator (DFIG) with a generation output of 500~MW is placed at the buses presented in Table~\ref{tab:cases}.

\begin{table}[h]
\caption{Representative buses to assess the impact of stochastic feed-in by IBRs.}
\label{tab:cases}
\centering
\begin{tabular}{cccc}
\hline \hline
\textbf{Bus} & \textbf{Nodal Inertia} [s]& \textbf{Fiedler eigenvector} & \textbf{GFV} \\ \hline
20           & 130.5                & 0.84              & 0.99  \\
51           & 983              & 0.71              & 0.60  \\
61           & 740            & 0.31            & 0.08 \\ 
53           & 537            & 0.34            & 0.02  \\\hline\hline
\end{tabular}
\end{table}

The Type-3 DFIG wind turbine (WT) is implemented in PST~\cite{Felipe2016}, utilizing a lumped model to represent a group of individual WT generators of the same type. It is assumed that the WT operates in the maximum power point tracking (MPPT) mode and not to contribute any virtual inertia. Consequently, the generator's speed decreases, resulting in a reduction in power output and, consequently, affecting the system frequency behavior.

The stochastic variations of wind speed are modeled using an Ornstein-Uhlenbeck (OU) process~\cite{Milano2013}. This process is given by the solution to the stochastic differential equation:
\begin{equation}
\label{eq:OUP}
\dot{\eta}(t) = \alpha(\mu - \eta(t)) + b\varepsilon,
\end{equation}
\noindent where $\eta(t)$ is the stochastic variable, $\alpha$ is the drift parameter, $b$ is the diffusion term, $\mu$ is the mean of the process, and $\varepsilon$ is the Gaussian white noise.  The parameters $\mu = 14$ m/s, $\alpha = 0.1$, and $b = 0.099$ were chosen. These values were selected to ensure that the wind speed variation has a variance of 5\% of $\mu$. For each Monte Carlo realization, a time step of 0.01s was chosen, consistent with the PST simulations, as recommended in~\cite{Milano2013}. The total of 200~s of simulation data are obtained each realization. To properly analyze the local and overall system response, for each realization, the frequencies of all buses and the COI frequency are recorded.

\subsubsection{Evaluation of the COI Frequency}

The COI frequency is used to analyze the impact of a local variation at a particular bus on the frequency behavior of all generators. Since the WT do not contribute with inertial response, the COI is calculated using the conventional formulation $\omega_{COI}=\sum_i H_i \omega_i/\sum_i H_i$, with $i$ related to the synchronous machines.

\begin{figure}[htb]
\includegraphics[width=2.9in]{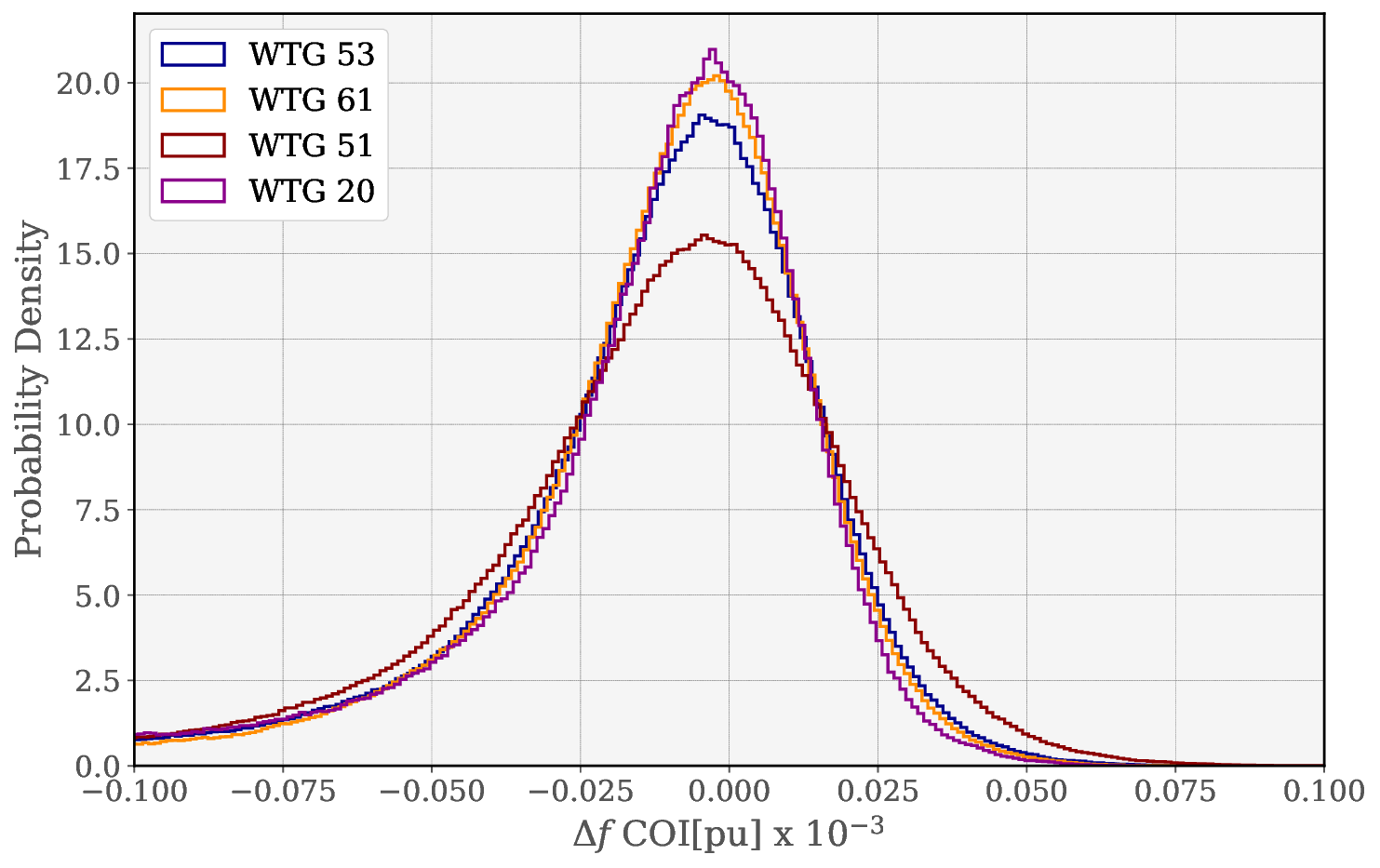}
\centering
\caption{Comparative of the COI frequency histogram for the four WTG placement cases.}
\label{fig:COIcomparative}
\end{figure}

Figure~\ref{fig:COIcomparative} illustrates the histogram depicting the frequency distribution of the COI for the four scenarios. Notably, the scenario with the least deviation in COI frequency occurs when the WTG is connected to Bus 20. This bus has the lowest nodal inertia and is located in a region that contributes significantly to the Fiedler mode. In contrast, the scenario with the most dispersed COI frequencies arises when the WTG is connected to Bus 51, which possesses the highest nodal inertia in the system and is situated in a region that strongly influences the Fiedler mode. This analysis indicates that buses with low nodal inertia demonstrate greater robustness against stochastic power injections. However, it is essential to recognize that the COI frequency reflects only the global behavior of the system, and thus, a local evaluation of individual bus frequencies and their impact on other buses in the system is necessary.

\subsubsection{Point of Interconnection (POI) Frequency}

\begin{figure}
\includegraphics[width=2.9in]{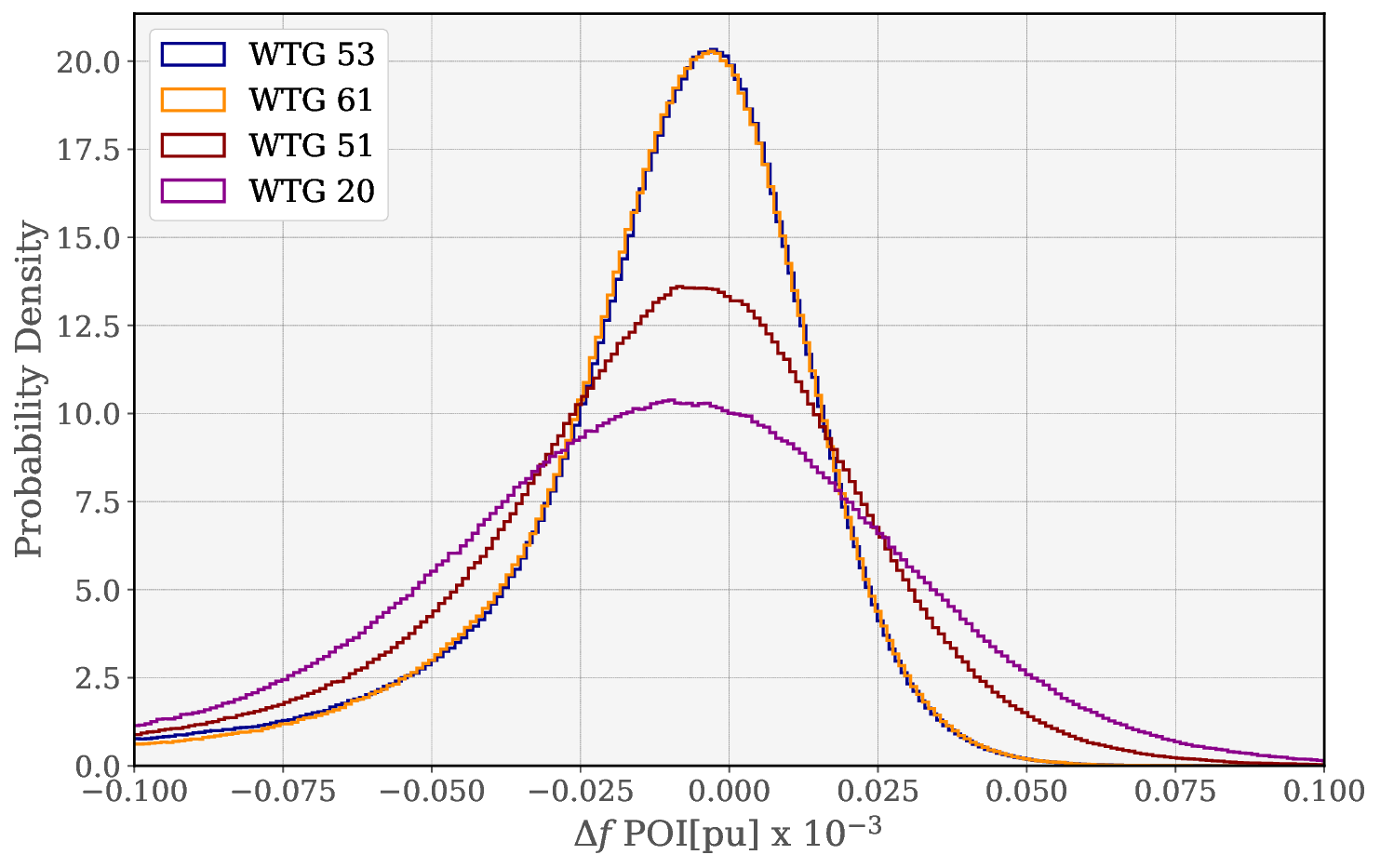}
\centering
\caption{Comparative of the histogram of POI frequency for the four cases of WTG placement.}
\label{fig:POIcomparative}
\end{figure}

To analyze the local frequency response, Figure~\ref{fig:POIcomparative} presents a histogram illustrating the frequency variation at the POI, specifically the local frequency of the buses where the WTG is located. In this case, Bus 20 exhibits the highest dispersion in local frequency, suggesting that this bus is more susceptible to local frequency variations resulting from stochastic generation. On the other hand, Buses 53 and 61 display similar behavior, indicating resistance to local frequency variations when subject to stochastic feed-in. 

Therefore, we can observe that nodal inertia plays a crucial role in both the POI and COI frequencies for buses with high values of the Fiedler eigenvector (see Figure~\ref{fig:all}(a) and Figure~\ref{fig:all}(b)). In these cases, the COI frequency response is better for buses with low nodal inertia, i.e., buses located farther from the COI. On the other hand, placing stochastic generation at buses with low nodal inertia results in more variation in the POI frequency. Furthermore, despite Bus 51 having the highest nodal inertia value, the buses with the least POI and COI frequency variations depend on both nodal inertia and the system topology (buses 61 and 53). Hence, it is evident that the proposed GFV metric effectively combines these two metrics to accurately indicate the region of the system that exhibits greater resistance to frequency variations induced by stochastic generation.
\vspace{-0.2cm}

\subsubsection{Disturbance Propagation Analysis}

\begin{figure}
\includegraphics[width=3.2in]{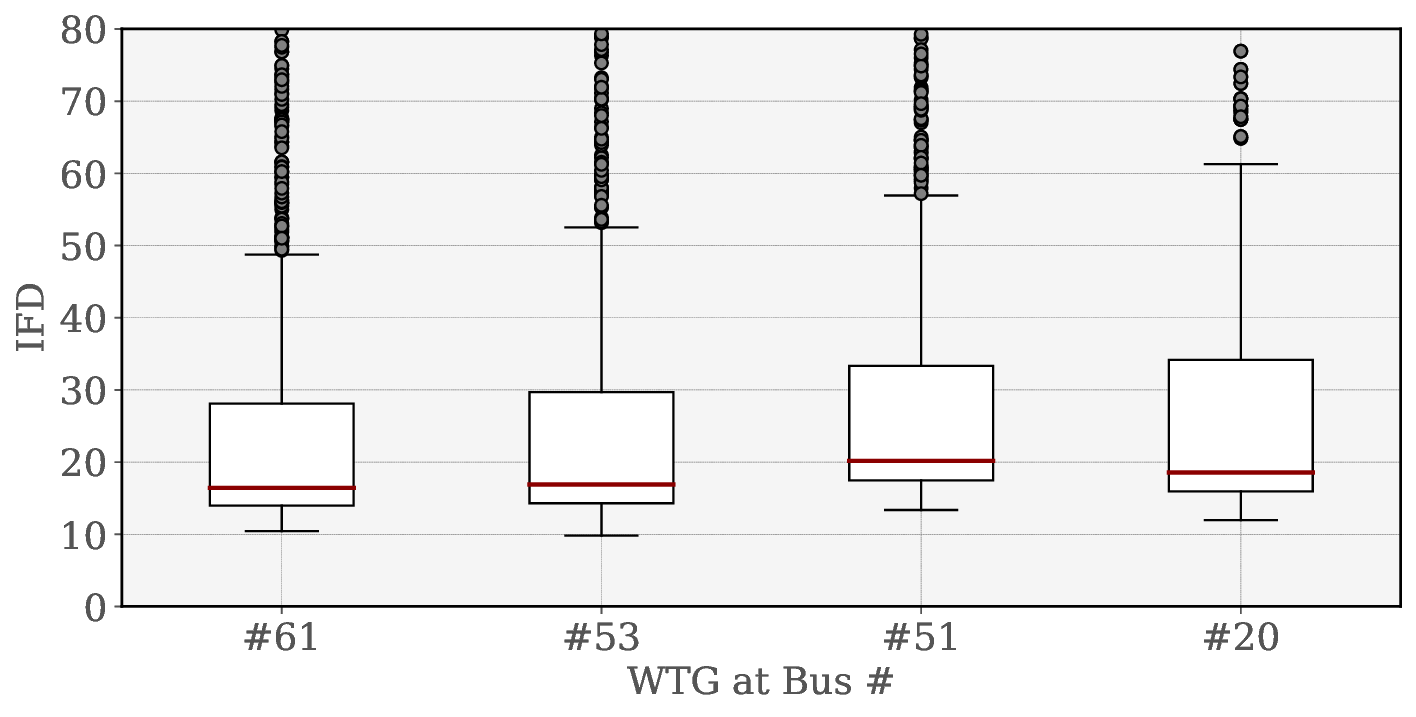}
\centering
\caption{Boxplot of the metric of IFD for the four cases of WTG placement.}
\label{fig:CIFDcomparative}
\end{figure}

To evaluate the spreading of frequency variations,  the integral frequency deviation (IFD) metric can be utilized~\cite{Zuo2021}:
\begin{equation}
\label{eq:IFB_eq}
\text{IFD} = \sum_{i=1}^{N_b}\sum_{k=1}^{N_s} |f_{k,i} - f_0|,
\end{equation}
\noindent where $f_{k,i}$ represents the frequency of bus $i$ at time step $k$, $N_b$ is the number of buses, $N_s$ is the number of samples, and $f_0$ is the nominal frequency, i.e., 1 pu. This metric is calculated for each simulation. Figure~\ref{fig:CIFDcomparative} displays the boxplots of the obtained IFD values for each of the aforementioned cases of WTG placement. Outliers are neglected in this analysis, focusing on the median (red line), inferior limit, and superior limits. It is observed that the case with the least variance in IFD values occurs when the stochastic feed-in is connected to buses 61 and 53. Conversely, the case with the highest variance in the IFD metric is when the WTG is connected to bus 20. Therefore, these results are in accordance with the proposed GFV, where buses 53 and 61 have the lowest values of GFV and demonstrate a better combination of local frequency response and lower impact on the COI response and disturbance propagation.

\section{Conclusion}

In this work, the concept of nodal inertia and spectral analysis of the network Laplacian matrix are used to derive a novel metric for evaluating nodal frequency strength. This metric can determine the locations most robust to frequency variations during stochastic IBR fluctuations. We demonstrate that the most robust buses, in terms of frequency strength, are those located in regions with lower contributions to the Fiedler mode and higher nodal inertia. From a practical perspective, the proposed GFV is particularly suited for offline planning and system analysis, where the optimal locations for integrating IBRs into the power system can be identified, as well as critical locations where the stochastic behavior of these resources could compromise frequency stability. Future research will explore the relationship of the proposed GFV metric with the dynamic flexibility \cite{Brahma2020} and overall systems robustness.

\vspace{-0.2cm}

\bibliographystyle{IEEEtran}
\bibliography{PSCC_2022.bib}

\end{document}